\documentclass[conference,10pt]{IEEEtran}
\IEEEoverridecommandlockouts
\usepackage{cite}
\usepackage{amsmath,amssymb,amsfonts}
\usepackage{algorithmic}
\usepackage{graphicx}
\usepackage{textcomp}
\usepackage{xcolor}
\usepackage{subfig}
\graphicspath{{figures/drawings/}{figures/plots/pdf/}}
\def\BibTeX{{\rm B\kern-.05em{\sc i\kern-.025em b}\kern-.08em T\kern-.1667em\lower.7ex\hbox{E}\kern-.125emX}}
\begin{document}

\title{ %
Combining Deep Learning and Linear Processing for Modulation Classification and Symbol Decoding 
}

\author{\IEEEauthorblockN{Samer Hanna\IEEEauthorrefmark{1}, Chris Dick\IEEEauthorrefmark{2}, Danijela Cabric\IEEEauthorrefmark{1}\thanks{This work was supported in part by the CONIX Research Center, one of six centers in JUMP, a Semiconductor Research Corporation (SRC) program sponsored by DARPA.}}\IEEEauthorblockA{\IEEEauthorrefmark{1}Electrical and Computer Engineering Department, University of California, Los Angeles}\IEEEauthorblockA{\IEEEauthorrefmark{2}Xilinx Inc., San Jose, California, USA}
\IEEEauthorblockA{samerhanna@ucla.edu, chrisd@xilinx.com, danijela@ee.ucla.edu}}

\maketitle

\begin{abstract}
Deep learning has been recently applied to many problems in wireless communications including modulation classification and symbol decoding. Many of the existing end-to-end learning approaches demonstrated robustness to signal distortions like frequency and timing errors, and outperformed classical signal processing techniques with sufficient training. However, deep learning approaches typically require hundreds of thousands of floating points operations for inference, which is orders of magnitude higher than classical signal processing approaches and thus do not scale well for long sequences. Additionally, they typically operate as a black box and without insight on how their final output was obtained, they can't be integrated with existing approaches. In this paper, we propose a novel neural network architecture that combines deep learning with linear signal processing typically done at the receiver to realize joint modulation classification and symbol recovery. The proposed method estimates signal parameters by learning and corrects signal distortions like carrier frequency offset and multipath fading by linear processing. Using this hybrid approach, we leverage the power of deep learning while retaining the efficiency of conventional receiver processing techniques for long sequences.  The proposed hybrid approach provides good accuracy in signal distortion estimation leading to promising results in terms of symbol error rate. For  modulation classification accuracy, it outperforms many  state of the art deep learning networks.
\end{abstract}


\section{Introduction}
Recently, deep learning was proposed to address many problems in wireless communications \cite{gunduz_machine_2019}. Deep learning has been used for identifying signal modulation~\cite{oshea_over--air_2018,west_deep_2017,hong_automatic_2017,icamcnet_2020}, estimating channels, and even building end-to-end communications~\cite{gunduz_machine_2019}. Deep learning approaches can be used to solve many problems where training data can be obtained and practical modeling based solutions are not tractable like  automatic modulation classification and blind symbol decoding.

Automatic modulation classification (AMC) is the problem of identifying the received signal type among a given set of modulations. Once, the modulation type has been recognized, blind symbol decoding aims to recover the transmitted symbols. These problems have many military and civilian applications.  Military applications would use AMC for the interception of hostile communications. In civilian applications, AMC could enable adaptive communications or facilitate communications between heterogeneous cooperating radios.

In the deep learning literature, modulation classification and symbol recovery have been addressed separately. For  modulation classification, many neural network architectures have been proposed and compared \cite{oshea_over--air_2018,west_deep_2017,hong_automatic_2017,icamcnet_2020}. Even though neural networks for modulation classification learn to be robust to distortions like noise and carrier frequency offset, the black box nature of deep learning does not enable the extraction of the necessary information for signal reconstruction. Some of the existing works have proposed using signal processing inspired layers to improve modulation classification~\cite{hanna_asilomar_2017,oshea_radio_2016,yashashwi_learnable_2018}, while others have used a dedicated network to estimate the distortions~\cite{oshea_learning_2017}. But, none has proposed an  efficient solution for both modulation classification and symbol recovery. Deep learning was also considered for decoding symbols of known signal types. Recurrent neural networks were proposed  to decode received symbols in an unknown communication channel \cite{farsad_neural_2018}. In \cite{ye_power_2018},  OFDM symbols were detected using neural networks. These approaches require a large number of FLOPS  compared to classical approaches and are designed under the assumption of a known   transmitted signal type.

 Works leveraging signal processing techniques have considered blind joint symbol recovery and modulation classification. However, they often make many simplifying assumptions, e.g. known frequency and timing offsets or channel~\cite{kazikli_optimal_2019}. In \cite{rebeiz_energy-efficient_2014}, a decision tree algorithm based on statistical tests for blind modulation classification and symbol recovery was proposed. Joint blind channel estimation, modulation classification, channel coding recognition, and data detection using an iterative algorithm was considered in \cite{liu_blind_2019}. One of the disadvantages of signal processing approaches is that they require a large number of samples for parameter estimation and modulation classification.

In this work, we propose a deep learning approach combined with receiver signal processing for joint modulation classification and symbol recovery. The proposed approach consists of two paths: a feature path based on neural networks and a signal path using linear  operations like filters. We refer to our approach as the Dual Path  Network (DPN). Both paths are connected by neural networks extracting features from the signal path and providing the parameters to restore the signal. The network incrementally reconstructs the signal and reuses it for a better estimation of parameters. The neural networks feature estimation and modulation classification require a very short sequence of input signal samples. The correction of input signal based on these parameters and decoding is performed using the linear signal path which can be applied efficiently on very long sequences.

The rest of the paper is organized as follows. The system model and the problem formulation are introduced in Section~\ref{sec:system}. The proposed Dual Path network is described in Section~\ref{sec:dpn}. In Section~\ref{sec:data}, we discuss datasets used in training and testing. The results are shown in Section~\ref{sec:results}. Section~\ref{sec:conclusion} concludes the paper.

\renewcommand{\b}[1]{\boldsymbol{\mathrm{#1}}}
\newcommand{\mC}{\mathbb{C}}
\section{System Model and Problem Formulation}
\label{sec:system}
A transmitter sends a vector of  complex symbols $\b{s} \in \mC^{N_s}$ using modulation type $M$ from a set of modulations $\mathcal{M}$. In the most general case, the transmitted signal $x(t)$ is determined by symbols $\b{s}$ and symbol duration $\tau$ through a modulation specific mapping function $\mathcal{G}$  such that
$x(t) = \mathcal{G}(\b{s},\tau)$.
 For a linear modulation type, the individual symbols $s_i$ represent a mapping from bits to a predefined constellation point, and the transmitted signal $x(t)$ given by 
$
	x(t) = \sum_{i=1}^{N_s} s_i p(t-i \tau)
$
where $p(t)$ is the pulse shaping filter. The signal is upconverted and  transmitted over a multipath fading channel modeled with an impulse response $h(t)$. The downconverted and sampled received signal is  modeled as the vector $\b{y} \in \mC^{N_r}$ 
\begin{equation}
y[k]= e^{j 2\pi( f_0 t_k + \phi_0)} \int_{-\infty}^{\infty} x(\sigma) h(t_k -\sigma) d\sigma  + n(t_k)
\end{equation}
where $f_0$ is the carrier frequency offset, $\phi_0$ the phase offset, and $n(t)$ is the additive white Gaussian noise. We assume the receiver sampling rate is $\tau_{0}$. Due to the sampling rate offset the sampling time $t_k$ is given by $t_{0} + k \tau_{0}$, where $\tau_{0}\geq \tau$ and $t_{0}$ the sampling phase offset such that $ 0 \leq t_{0} \leq \tau_0/2$. 
The length of transmitted and received symbols is related as
$N_r = N_s \left\lceil \frac{\tau_0}{\tau} \right\rceil$.

Given vector $\b{y}$, the receiver's objective is to identify the modulation type $M$ and recover the transmitted symbols~$\b{s}$. The signal identification  should be accurate using short sequences and  the recovery scalable to  long sequences in a computationally efficient manner. 

\section{Dual Path Network (DPN)}
\label{sec:dpn}
\begin{table}
	\renewcommand{\arraystretch}{1.5}
	\caption{Output description \label{tbl:op_desc}}
	\centering{
		\begin{tabular}{|c|c|c|c|}
			\hline
			Name & Description & Equation \\	\hline
			Op1 & Noise removed &	$z_1[k] = y[k] - n(t_0 + k \tau_0)$ \\ \hline
			Op2& Frequency corrected &	$z_2[k] = e^{-j 2\pi f_0 (t_{0} + k \tau_{0})} z_1[k]$ \\ \hline
			Op3& Recovered Signal &	$z_3[k] = x(t_{0} + k \tau_{0})$ \\ \hline
			Op4& Timing information &	$z_4[k] = g_t(t_{0} + k \tau_{0} )$ \\ \hline
			Op5& Modulation type &	$z_5 = g_m(M)$ \\ \hline
		\end{tabular} 
	}
\end{table}
\begin{figure*}[htbp]
	\centerline{\includegraphics[scale=1]{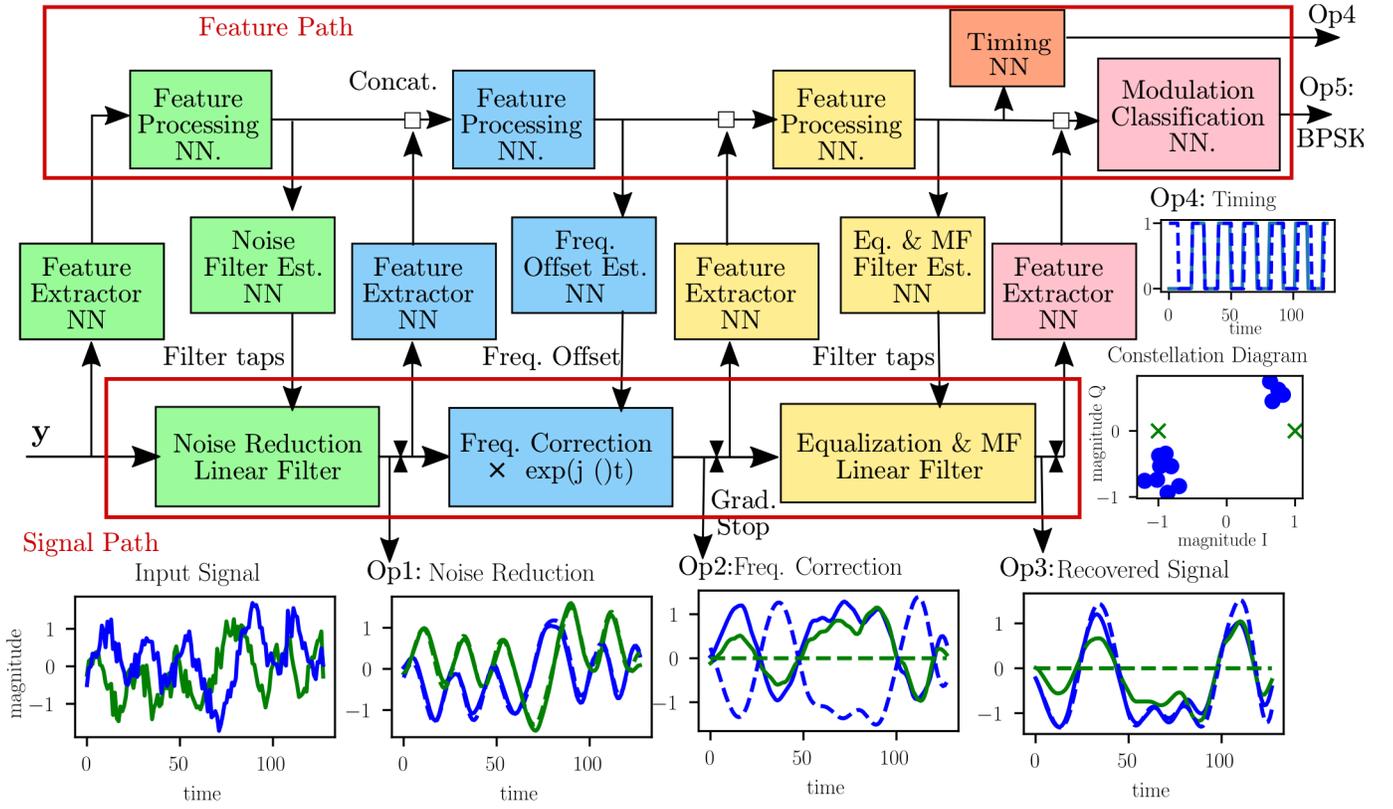}}
	\caption{The Dual Path network consists of feature  path and a linear signal path connected using neural networks (NN) for parameter estimation and feature extraction. An example input signal is shown   along with the predictions in solid and the reference output in dashed. The output constellation is obtained by sampling op3 using op4. }
	\label{fig:dualPath}
\end{figure*}
The proposed network architecture is inspired by the signal demodulation flow used in conventional digital demodulators when the modulation type, pulse shape, symbol rate, and carrier frequency are known \emph{a priori}. In a typical demodulation flow, any residual errors or offsets due to the lack of synchronization are estimated and corrected one after the other~\cite{prasad_lets_2011,hanna_maximizing_2016}. The compensation of these errors is typically implemented using linear operations like filters. Under the lack of knowledge of the transmitted signal type and parameters, the classical demodulation approaches are not applicable. To identify the signal and estimate its parameters, we explore using deep learning. Deep learning relies on the availability of training data making it easy to apply to unknown signals.

The proposed network consists of two paths: a signal path consisting of linear operations inspired by existing signal processing methods, and a feature path where deep neural networks (NN) learn different signal parameters. The overall network is shown in Fig.~\ref{fig:dualPath}.  Both paths are connected using a set of neural networks. Feature extractor NNs process the signal to learn features. Parameter estimator NNs use the learned features to estimate the parameters and feed them to the signal path for correction and reconstruction of the signal. 
As in a typical demodulation flow, the signal reconstruction and parameter estimation are done incrementally. We start with noise estimation and reduction, followed by correction of frequency offset, matched filtering and equalization. Using this incremental approach, each stage benefits from the correction performed by the previous stage.%

\subsection{Architecture}
\begin{figure}[htbp]
	\centerline{\includegraphics[scale = 1]{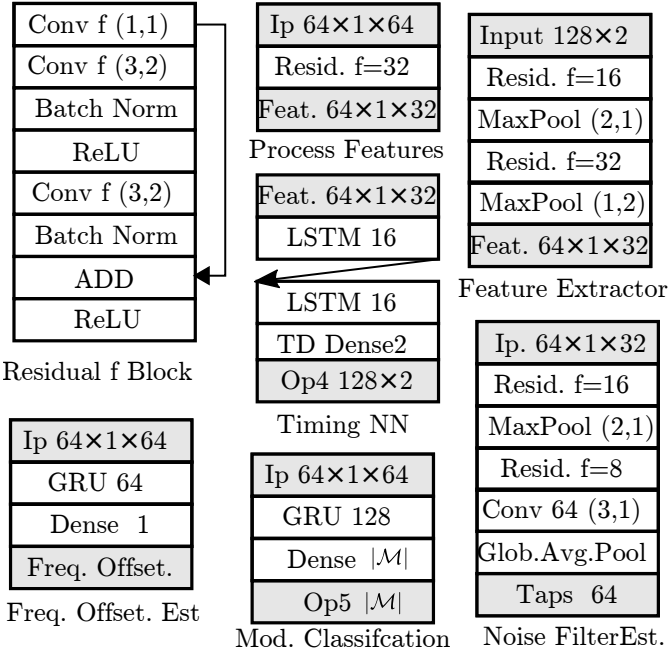}}
	\caption{The layer by layer description of the NN in DPN.}
	\label{fig:dualPathLayers}
\end{figure}

The network takes one input, which is the received samples $\b{y}$, and generates five outputs (Op) as shown in Table \ref{tbl:op_desc}. An example for a BPSK signal is shown in Fig.~\ref{fig:dualPath}. The first three outputs are processed signals with  distortions correction, namely noise reduction, carrier frequency offset correction, and equalization. The fourth output  estimates  the timing errors and specifies the ideal sampling time. The function $g_t$ generates a binary vector with the same length as the signal having transitions at the  sampling time as shown in the timing plot in Fig.~\ref{fig:dualPath}. The last output $z_5$ is the one-hot encoding given by the function $g_m$ of the modulation type.

The neural network (NN) structure of each block is shown in Fig.~\ref{fig:dualPathLayers}. The entire network consists of a combination of residual blocks and recurrent neural networks. Residual blocks improve the gradient flow and enable the training of deep networks~\cite{he_deep_2015}. They were also shown to give superior performance for modulation classification~\cite{oshea_over--air_2018}.   The ``Equalization (Eq.) and Matched Filter (MF) Estimator" is similar to the "Noise Filter Estimation" block shown in Fig.~\ref{fig:dualPathLayers} except that the former has 65 filter taps and the latter 64. Both estimation NNs consist entirely of convolutional layers and global average pooling was used to generate the filter taps.   The information contained in the timing signal (Op4)  consists only of the symbol rate and the timing phase. Both unknowns need to be estimated from the entire input sequence. Based on that the ``Timing NN" was designed as two LSTMs passing only the internal state. The first one scans the entire sequence and passes to the following LSTM its low dimension  internal state. The second network uses this output to generate the output sequence. A  time distributed dense network with sigmoid activation is later used to provide the required output format.
As for the signal path, the ``Noise reduction" and ``Equalization \& MF" are implemented as linear filters performing convolution using the estimated filter taps. Note that unlike  the convolutional layer in neural networks that operates on the fixed trained  weights and an intermediate output, the signal path convolution is performed between two intermediate  outputs from prior layers.
The ``Frequency Correction" performs complex multiplication with a complex exponential using the estimated frequency offset. As for the sizes of the hidden layers, they were decided based on a tradeoff between the network size and performance. Several values for these dimensions were evaluated before choosing the ones  in Fig.~\ref{fig:dualPathLayers}.

As stated earlier, the signal path consists  exclusively of linear operations. While this design choice sacrifices the ability to correct for nonlinear distortions, it brings many benefits; first, for long sequences, once the modulation type, timing information, frequency offset, and filter taps have been estimated, there is no need to keep inferring them using the neural network. Typically  neural networks used in signal processing consist of hundreds of thousands of parameters, and inference requiring hundreds of thousands of  floating point operations. Using this design, the inferred parameters can be reused and applied to very long sequences using simple operations. Second, the estimated parameters are interpretable and compatible with existing signal processing approaches. For example, if the frequency offset is variable during the signal duration, a phase locked loop can be used to track it.

\begin{table}
	\renewcommand{\arraystretch}{1.5}
	\caption{Loss Functions \label{tbl:loss}}
	\centering
	\begin{tabular}{|c|c|}
		\hline
		Equation \\	\hline
		$L_1 = \frac{1}{N_r}  (\|\hat{\b{z}}_1 - \b{z}_1\|^2) $ \\ \hline
		$L_2 =\frac{1}{N_r} (\hat{\b{z}}_2^H\hat{\b{z}}_2  + \b{z}^H_2 \b{z}_2 - 2 |\hat{\b{z}}_2^H  \b{z}_2| )$ \\ \hline
		$L_3 =\frac{1}{N_r}  (S(\hat{\b{z}}_3^H)S(\hat{\b{z}}_3)  + S(\b{z}_3^H) S(\b{z}_3) - 2 |S(\hat{\b{z}}_3^H)  S(\b{z}_3)| )$ \\ \hline
		$L_4 = \min\{  \mathcal{L}_{bv}(\b{z}_4,\hat{\b{z}}_4), \mathcal{L}_{bv}(1-\b{z}_4,\hat{\b{z}}_4) \} $ \\ \hline
		$L_5 =\mathcal{L}_{c}(\b{z}_5,\hat{\b{z}}_5) $ \\ \hline
		$L =\sum_{i=1}^{5} w_i L_i $ \\ \hline
	\end{tabular} 
\end{table}
\subsection{Training}
Typically neural networks rely on non-linear operations between layers for training. Since the signal path is designed to have linear operations, the  signal outputs (Op1, Op2, Op3) are necessary for each stage to perform its task.   Additionally, gradients are prevented from backpropagating into the signal processing blocks from the following layer since  the desired output is already provided. During training, additive white Gaussian noise is added to the signal before the "Equalization \& MF" stage to improve its training in the high SNR regime. 

Since this network has multiple outputs, the training loss is a combination of different losses shown in Table \ref{tbl:loss}. The  network output for $\b{z}_i$ is given by $\hat{\b{z}}_i$. For Op1, to reduce the noise, we use the mean squared error loss. For Op2, we use a loss function that does not penalize constant phase shift between vectors, such that $\hat{\b{z}}_2$ and phase shifted  $\hat{\b{z}}_2 e^{j\phi}$, where $\phi$ is the phase shift, would yield the same loss value. The rationale behind this choice is to handle phase ambiguity; if we consider a BPSK signal $\b{x}$, the signal $-\b{x}$ is also a valid BPSK signal and both can not be distinguished without side information. For  Op3, we use the same phase insensitive  function but only apply it to a downsampled version of the signal. Since the signal was oversampled, to get the constellation we are only interested in the downsampled version.  To that end, we define the sampling vector function $S(\cdot)$ such that
\begin{equation}
[S(\b{x})]_i = 
\begin{cases}
x[i] &  z_4[i]\neq z_4[i+1]\\
 0 & \text{Otherwise}
\end{cases}
\end{equation}  
This  function only considers the  sampling instances, which occur  at the transitions of the timing vector $\b{z}_4$ (see Op4 in Fig.~\ref{fig:dualPath}). For the "Timing NN" output Op4, we apply a vector binary crossentropy loss, such that  $\mathcal{L}_{bv}(\b{x},\b{y})= \frac{1}{N_r} \sum_{1}^{N_r} \mathcal{L}_b(x[i],y[i]) $ where $\mathcal{L}_b$ is the  binary crossentropy loss. Since the information lies in the transition and not the values, we consider the minimum loss of $z_4$ and its inverse $1-z_4$. As for "Modulation Classification" NN, we use a categorical crossentropy loss $\mathcal{L}_c$. The total loss is a weighted combination of these losses with weight vector~$\b{w}$. DPN was implemented using the KERAS API of TensorFlow. The optimizer used for training is the ADAM optimizer with a learning rate of 0.001  and the gradients were clipped at a norm of 1.0. The entire network was trained simultaneously. This method of training makes the NN of each stage adapt to the statistics of the previous output.

\section{Data Generation and Datasets}
\label{sec:data}
\begin{figure}[t]
	\centerline{\includegraphics[scale=1]{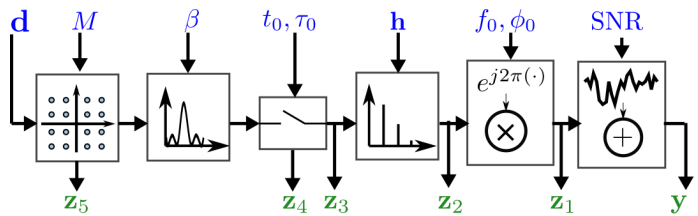}}
	\caption{Flow graph for generating samples showing on top the input parameters and the bottom the outputs used for training.}
	\label{fig:generator}
\end{figure}
We generate datasets consisting of samples with different data, modulation types, symbol rates, timing and frequency offsets, phase, channel impulses, and SNRs. The datasets emulate signals with unknown parameters being intercepted using a coarse frequency estimate and oversampling. Each sample is generated according  to the flow graph shown in Fig.~\ref{fig:generator}. Random data $\b{d}$ is generated and modulated using modulation type $M$ selected from the set of modulation $\mathcal{M}$. If $M$ is a linear modulation, the output is pulse shaped with a root-raised-cosine filter with a roll-off factor $\beta$. The output is sampled with an offset $t_0$ and a sampling time $\tau_0$. Multipath fading is simulated using convolution with random fading taps  having a delay spread~$\sigma$. Then frequency and phase offsets, $f_0$ and $\phi_0$, are applied, and Gaussian noise is added to model different SNRs.

 All aforementioned signal parameters are chosen randomly from  specified ranges. Two datasets are considered with $N_r=128$ and each dataset is defined by the range of each parameter as given by Table \ref{tbl:datasets}. Both datasets have $\beta \in \{0.15,0.35,.55\}$, $t_0 \in [0,\tau_0/2]$, $\phi_0\in [0,2\pi]$, $\b{h}$ has  3   non zero taps  having  $\sigma \in [0.5 \tau/\tau_0,4\tau/\tau_0] $ with the non-line-of-sight taps  having  average magnitudes of 0.5 and 0.1.  Dataset 1 has fewer modulations and less severe distortions, %
   while Dataset 2 is more challenging due to more modulation types, larger frequency offsets, and significantly different values of samples per symbols $\tau_0/\tau$. Dataset 2 is used in the  evaluation of the signal and symbol recovery.
 
Typically, a fixed dataset is used in training,  and data augmentation is performed to avoid overfitting. Since our dataset is generated using simulation, instead of fixing the training data, we generate the samples in real-time during training. This means that each epoch consists of a new set of samples which effectively eliminates overfitting. As for validation and testing, two fixed datasets are used with one million samples in each.

\begin{table}
	\renewcommand{\arraystretch}{1.5}
	\caption{Dataset Description \label{tbl:datasets}}
	\centering
	\begin{tabular}{|c|p{0.75 in}|p{1.7 in}|}
		\hline
		Param. & Dataset 1 & Dataset 2 \\	\hline
		$M$ &\{ BPSK, QPSK, PSK8, QAM16, QAM64, GMSK, CPFSK, ASK4 \} &	\{OOK, ASK4, ASK8,  
		BPSK, QPSK, PSK18, PSK16, PSK32, APSK16, APSK32, APSK64, APSK128, QAM16, QAM32, QAM128, QAM256, GMSK, CPFSK\} \\ \hline
		$f_0 (Hz)$& $[0,0.0025/\tau_0]$ &	$[0,0.005/\tau_0]$ \\ \hline
		SNR (dB) & $[-20,20]$ &	$[-10,40]$ \\ \hline
		$\frac{\tau_0}{\tau}$ & $[7,9]$ &	$[3,16]$ \\ \hline
	\end{tabular} 
\end{table}

\section{Results}
\label{sec:results}
\subsection{Modulation Classification}
\begin{figure*}[t!]
	\begin{center}
		\subfloat[Comparison on dataset 1 \label{fig:radioml_bench}]{\includegraphics{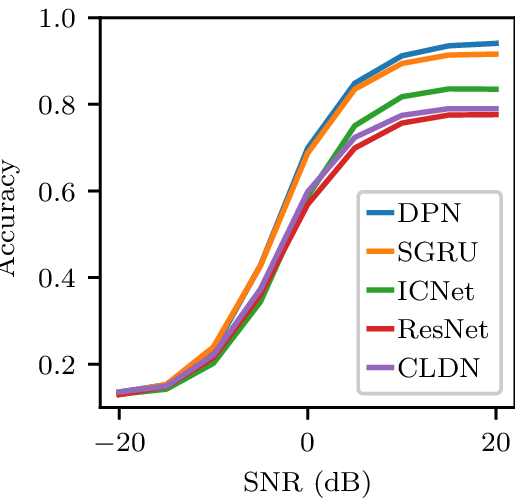}}  \hspace{1mm}
		\subfloat[Comparison on dataset 2 \label{fig:gen2_bench}]{\includegraphics{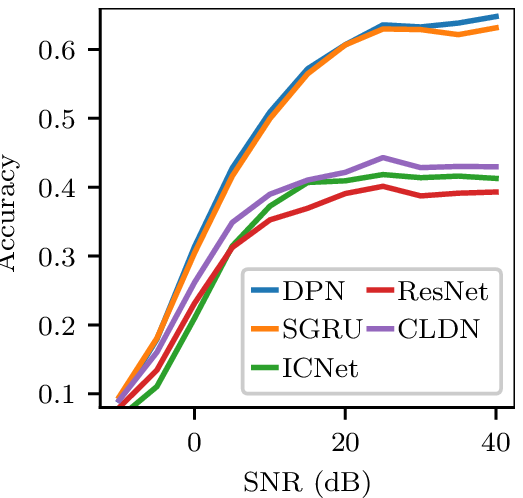}}  \hspace{1mm}
		\subfloat[A breakdown of  DPN on dataset 2 \label{fig:dualPath_arch}]{\includegraphics{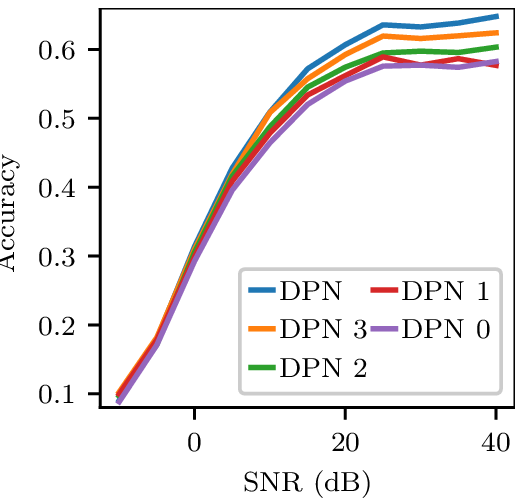}}  \hspace{1mm}
	\end{center}
	\caption{Modulation Classification Results.}
	\label{fig:mod}
\end{figure*}

\begin{figure}[t]
	\centerline{\includegraphics[scale=1]{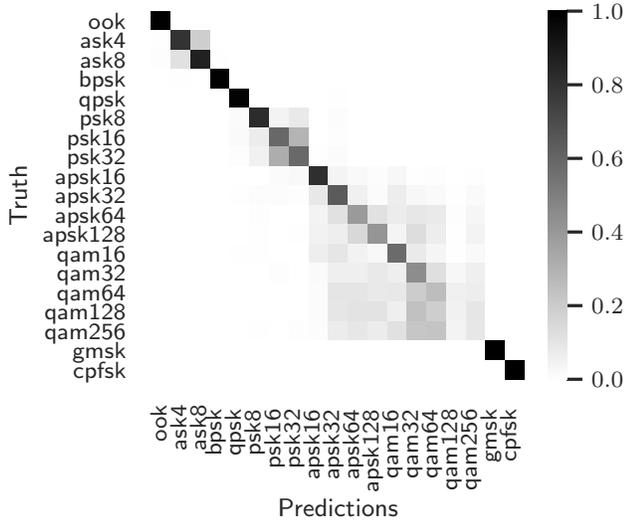}}
	\caption{Confusion matrix of DPN at SNR=40dB. Due to the short sequence length, misclassification occurs between high order modulations.}
	\label{fig:confusion}
\end{figure}
We evaluate the modulation classification performance of our proposed DPN and compare it to the state of the art approaches for modulation classification. Namely, we consider the ResNet architecture~\cite{oshea_over--air_2018}, the CLDNN architecture~\cite{west_deep_2017}, the Stacked GRUs (SGRU)~\cite{hong_automatic_2017}, and ICNet~\cite{icamcnet_2020}. All these approaches use as input IQ samples and  directly predict the modulation class without generating any other information about the signal. In terms of the number of parameters, DPN has 189K trainable parameter  which is about the same number as the smallest network. 

For Dataset 1, DPN was allowed up to 100 epochs, and for Dataset 2 DPN had up to 200 epochs. Since DPN has access to the four intermediate stages of the signal (Op1 to Op4), to be fair in comparison the remaining networks were allowed to have up to 4 times more data and training epochs.
 Each epoch consists of 800K samples and the batch size was adjusted for maximum GPU utilization.   The network training was stopped if the validation loss did not improve for ten epochs. The results for Dataset 1 is shown in Fig.~\ref{fig:radioml_bench} and for Dataset 2 in Fig.~\ref{fig:gen2_bench}. From these figures, we see that DPN significantly outperforms most of the existing approaches except for the SGRU.  The SGRU performs close to DPN but does not provide symbol decoding. Hence, DPN performs as good or better than the state of the art approaches in modulation classification. 
 
 The performance on Dataset 2 is lower than Dataset 1 and does not exceed 65\% for any of the approaches. By looking at the confusion matrix at 40dB SNR in Fig.~\ref{fig:confusion}, we can see that this performance is attributed to errors in high order modulations.   This is expected since we consider high order modulations up to QAM 256. Also, Dataset 2 has samples  up to 16 samples per symbol.  For a sequence of 128 samples, this means that each sequence can have as little as 8 symbols, which makes it difficult to distinguish high order modulations. 

To understand the significance of having the intermediate  signals in DPN, we train several partial instances of DPN. In all these instances, the feature path is the same and we incrementally add the signal stages and the corresponding extractors and estimators. DPN 0 is obtained by removing the signal path  and the timing module, hence, the network is trained similar to the existing approaches. For DPN 1 we add the noise filtering stage. For DPN 2, we add the first two stages, and for DPN 3 we add all signal stages. Our original DPN contains all 3 signal stages and the timing module. Fig.~\ref{fig:dualPath_arch} shows that each stage incrementally improves the performance of modulation classification.

\subsection{Parameter Estimation}
\begin{figure*}[htbp]
	\begin{center}
		\subfloat[SNR \label{fig:eval_snr}]{\includegraphics[scale=0.95]{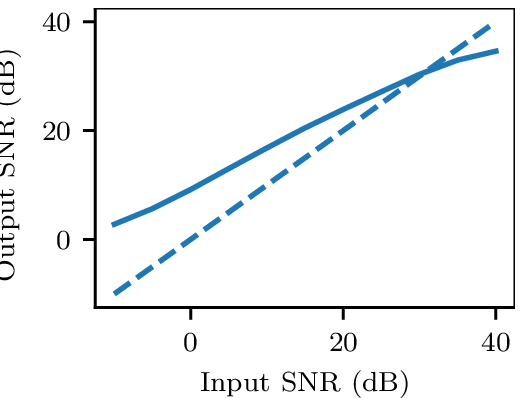}}  \hspace{1mm}
		\subfloat[Frequency Offset \label{fig:eval_freq}]{\includegraphics[scale=0.95]{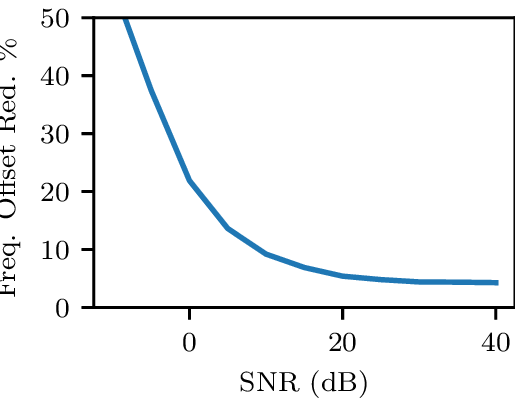}}  \hspace{1mm}
		\subfloat[Timing Error \label{fig:eval_timing}]{\includegraphics[scale=0.95]{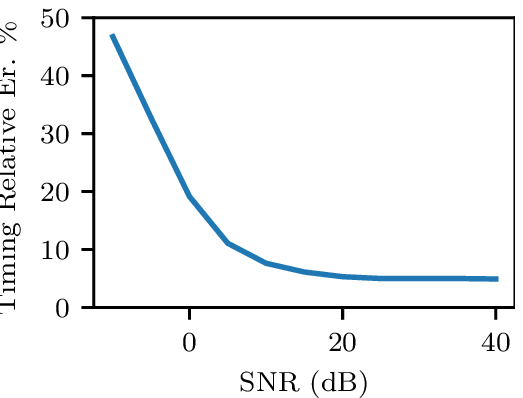}}  \hspace{1mm}
	\end{center}
	\caption{Performance of DPN in parameter estimation.}
	\label{fig:params}
\end{figure*}

The performance of DPN in terms of parameter estimation is evaluated on the signals in the test set of Dataset 2 and the averaged results are shown in Fig.~\ref{fig:params}. Fig.~\ref{fig:eval_snr} shows the SNR of the predicted  signal $\hat{\b{z}}_1$ plotted against the SNR of the input signal $\b{y}$. We see that the noise reduction stage significantly increases the SNR for low SNR signals. For very high SNR signals, above 30 dB, the first stage seems to add small amounts of noise to the signal. However, for high SNRs, this loss does not have any significance for the symbol recovery. To evaluate the improvement in frequency estimation, we calculate the ratio between the residual frequency offset after correction and before correction $\frac{ \mathbb{E} \{f_0 - \hat{f}_0\}}{\mathbb{E} f_0}$ where $\mathbb{E}$ is the mean calculated per SNR and $\hat{f}_0$ is the estimated offset. This improvement is shown in Fig.~\ref{fig:eval_freq}. We can see that for high SNR, the carrier frequency offset gets reduced to below $5\%$. For symbol rate estimation,  the average absolute error per SNR given by $\mathbb{E}\frac{|\tau - \hat{\tau}_0|}{ \tau_0}$  is shown in  Fig.~\ref{fig:eval_timing}. Again, DPN achieves a low timing estimation error for SNR above 5dB.  It is worth noting that these estimates are obtained from a very short signal consisting of 128 samples without knowing the signal type.
\subsection{Symbol Recovery}
\begin{figure}[t]
	\begin{center}
		\subfloat[PSK \label{fig:ser_0}]{\includegraphics{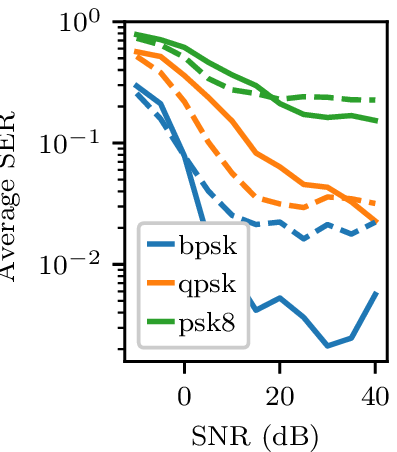}}  \hspace{1mm}
		\subfloat[QAM \label{fig:ser_1}]{\includegraphics{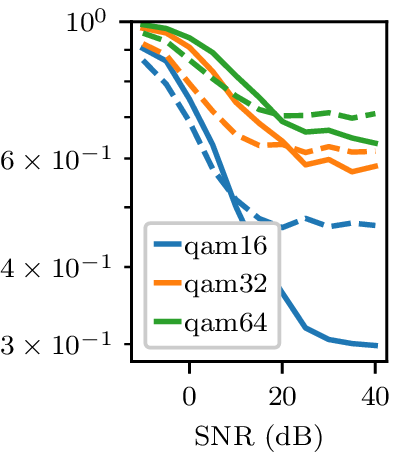}}  
	\end{center}
	\caption{SER of DPN in solid is compared against the reference DSP approach in dashed for a subset of modulations.}
	\label{fig:ser}
\end{figure}
We  evaluate DPN's ability to decode the symbols blindly. Since the received signal is short, distorted, and with unknown paramters, we expect to achieve relatively high symbol error rate (SER) even at high SNRs. To evaluate the performance of DPN, we use a simple signal processing approach to provide a baseline reference. In this evaluation, we focus on the  symbol error rate, and to that end, for both approaches we make the following assumptions: (1)  modulation type was inferred correctly and this is valid for low order modulations; (2) the sampling instances were accurately determined to make sure that the compared symbols are aligned; (3) the phase is accurately recovered to handle phase ambiguity. For both approaches, we only consider symbol recovery for linear modulations and use the conventional minimum Euclidean distance receiver.

 Note that for many of the classical estimation approaches, a vector of length 128 is too short to derive an accurate estimate of the signal parameters. Therefore, as a reference signal processing approach, we assume a genie approach for the frequency recovery, and we consider a fixed low pass filter that works for all samples in the dataset. Although, there exists methods for blind channel equalization, they are typically slow to converge and require a known modulation type~\cite{ahmed_review_2019}. Due to the short sequence and the lack of channel state information, no channel equalization is performed in the DSP approach. 
 
 The results for the symbol error rates (SER) at different SNRs for PSK and QAM modulations are shown in Fig.~\ref{fig:ser_0} and Fig.~\ref{fig:ser_1}. We can see that at high SNR, DPN tends to outperform the DSP approach. We explain this result by the fact that DPN learns to perform blind equalization on the short input signal, which reduces the SER. At the lower SNRs, DPN frequency correction is not very good. This leads to more distortion to the signal leading to high SER. These results show that the deep learning DPN is able to match the performance of the used DSP approach and even surpass it. However, for both approaches the SER is relatively high even at high SNR. This is expected given that short signals with no preamble and unknown parameters are used.
 
\subsection{Remarks on DPN Complexity and Scalability}
\begin{figure}[t]
	\centerline{\includegraphics{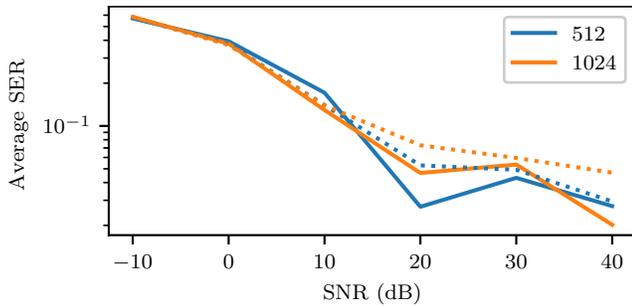}}
	\caption{Comparison of SER for QPSK when reusing parameters in solid and when re-estimating parameters in dotted. There is no significant difference in performance.}
	\label{fig:bg_pkt}
\end{figure}
One of the advantages of the Dual Path network is its ability to perform symbol recovery over long sequences efficiently. In contrast, fully NN based approaches would have to be applied over long sequences at a high computational cost. For an input of 128 samples, DPN consumes about 383 KFLOPS, out of which signal path only uses 99KFLOPS. That's about 4 times less operations (without combining linear operations).

The FLOP count for the signal path was approximately estimated as follows: each of the 2 convolution stages consumes about 128*64 complex operations  and the frequency correction consumes 128 complex operations. This sums to 16512 complex operations. With each complex multiplication containing 6 real operations, this translates to 99KFLOPS. For the neural networks, the TensorFlow profiler estimated the FLOP count to 283K. 

 To evaluate the effect of reusing the estimated parameters on SER, we generate 1000 QPSK signals of lengths  512, and 1024. These signals are divided into chunks of 128.  We compare the SER results when DPN is applied on all chunks to the case when it is applied to the first chunk and then the signal path  reuses the estimated parameters on the remaining chunks. The results in Fig. \ref{fig:bg_pkt} show that there is no significant  impact from reusing the parameters. Hence, we can get about a 4 times reduction in FLOPS to achieve the same SER for subsequent chunks.

\section{Conclusion and Future Work}
\label{sec:conclusion}
We have proposed the Dual Path Network for joint blind modulation classification and symbol recovery. By combining neural networks along with linear signal processing, we can leverage the power of deep learning for parameter estimation while retaining the efficiency of classical signal processing techniques for long sequences without sacrificing performance.  Results show that DPN can estimate the signal parameters for SNRs above 5dB with a very low number of samples and that the performance of modulation classification surpasses many of the state of the art networks. The successful reconstruction enables blind symbol recovery with symbol error rates lower than a genie signal processing approach without equalization under high SNR. 
In our future work, we will further analyze the performance of each stage individually. The effect of  sequence length on the overall performance will be studied. We will also compare our approach with a non genie, fully blind signal processing approach for symbol recovery and parameter estimation.

\bibliographystyle{IEEEtran}
\bibliography{references}

\end{document}